# Central Diffraction at ALICE

Jerry W. Lämsä[1,2] and Risto Orava[1]

[1]Helsinki Insitute of Physics,  and Division of Elementary Particle Physics, Department of Physics,
PL 64 (Gustaf Hällströminkatu 2a), FI-00014 University of Helsinki, Finland
[2]Physics Department, Iowa State University,  Ames, 50011 Iowa, U.S.A.

## Abstract

The ALICE experiment is shown to be well suited for studies of exclusive final states from central diffractive reactions. The gluon-rich environment of the central system allows detailed QCD studies and searches for exotic meson states, such as glueballs, hybrids and new charmonium-like states. It would also provide a good testing ground for detailed studies of heavy quarkonia. Due to its central barrel performance, ALICE can accurately measure the low-mass central systems with good purity. The efficiency of the Forward Multiplicity Detector (FMD) and the Forward Shower Counter (FSC) system for detecting rapidity gaps is shown to be adequate for the proposed studies. With this detector arrangement, valuable new data can be obtained by tagging central diffractive processes.

# 1 Introduction

An efficient central diffraction physics program can be realized with the ALICE experiment[1]. Low-mass central diffractive final states decaying into a small number of particles, with no forward secondary (shower) particles produced from interactions in the beam pipes, can be selected with the Forward Multiplicity Detectors (FMDs) and Forward Shower Counters (FSCs) [1,2] on both sides of the ALICE experiment at the LHC Interaction Point (IP2).

The central diffractive (CD) reaction

$$p\ p \rightarrow p + M + p, \qquad (1)$$

where M is a hadronic state, is selected with the use of FMDs and FSCs, which define the rapidity-gaps (denoted +).[2] Detection of the outgoing protons, which will mostly remain within the beam pipe, is not required for this study. The reaction is considered to be mediated by the two-gluon colour-singlet interaction

$$g\ g \rightarrow M. \qquad (2)$$

The physics program will be primarily concerned with a search for the production of meson states such as glueballs, hybrids, and heavy quarkonia $\chi_c$, $\chi_b$ [2,3]. A search will be made for structure in the mass spectra of exclusive decay states of M, such as, $\pi^+\pi^-$, $K^+K^-$, $2\pi^+2\pi^-$ and $K^+K^-\pi^+\pi^-$, $K^+K^o\pi^-$, $K^-K^o\pi^+$, $p\bar{p}$, $\Lambda\bar{\Lambda}$ and others [4]. A strong coupling for the reaction $g\ g \rightarrow M$ is expected as a result of the two-gluon exchange. Being central to QCD, discovery of glueballs would be of great importance. The very high statistics studies of the process in Eq. 2 provides this possibility.

For the case of pomeron – pomeron interactions, the central system is dominantly produced with spin-parity $J^{PC} = 0^{++}$, $2^{++}$, etc. The decays with low multiplicities such as $\pi^+\pi^-$ or $K^+K^-$ can be used as efficient spin-parity analyzers [2]. The t-channel exchanges over the large rapidity gaps can only be colour singlets with Q = 0. Known exchanges are the photon $\gamma$ and the pomeron P.[3] Another possible, but not yet observed, exchange in QCD is the odderon, O, a negative C-parity partner to the P with at least 3 gluons. The physics programme includes sensitivity to odderon exchange. Double pomeron exchange (DPE) produces primarily $I^G J^{PC} = 0^+ 0^{++}$ states, with some $0^+ 2^{++}$ admixture. $J^{PC} = 1^{--}$ states such as J/ψ and Y are produced by γP, but can also be produced by OP.

---

[1] The early LHC program, with anticipated 5 TeV + 5 TeV proton – proton interactions, at a low luminosity (no pile-up), is considered in this study.
[2] Concerning the purity of the exlusive central diffractive process, an earlier study by the authors is referenced, see [1].
[3] The gluon passes the Q requirement, but is not a colour singlet; however one or more additional gluons can cancel its colour and form a Pomeron.

Other interesting reactions involving heavy-flavor pairs should also be considered, for example, $D^o$, $\overline{D^0}$ and $B^o$, $\overline{B^o}$ which would be produced in even CP states.
The gluon-rich and quantum number filtered central system is a laboratory for studies of QCD and glueball spectroscopy. Studies on production rates of $\eta$, $\eta'$ mesons, baryons, etc. could be compared with inclusive $p\overline{p}$ interactions at $\sqrt{s} \approx M$, where M is the mass of the central system [3].

Based on forward shower counters, the CDF experiment at the Tevatron has recently produced a series of highly valuable results on central exclusive production of $\chi_c$, $\gamma\gamma$, di-jets and $J/\psi$ [6].

## 2 Experimental Overview

The Forward Multiplicity Detectors (FMD) cover the pseudorapidity range $-3.7 < \eta < -1.7$ and $1.7 < \eta < 5.0$ [2]. To extend the rapidity gap detection, FSC scintillation counters would be employed surrounding the beam pipes in the region from 20m to 100m on both sides of the LHC interaction point (IP2), see Fig.1. The FSCs will detect showers from very forward particles interacting in the beam pipe and surrounding material. The absence of a shower indicates a rapidity gap.

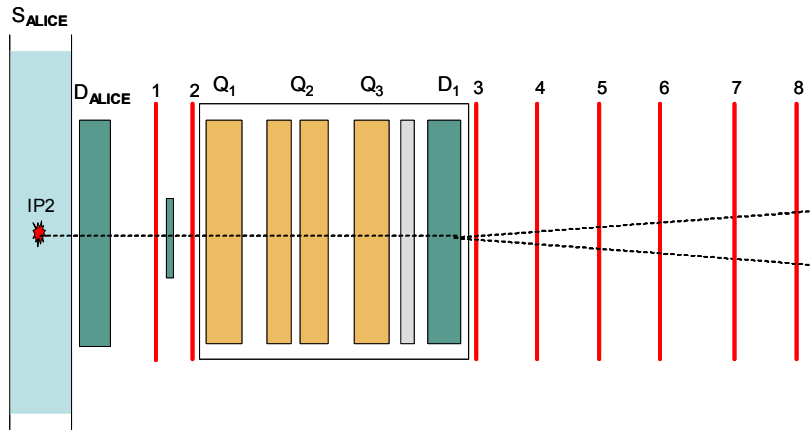

**Figure 1.** *The layout of ALICE detectors at the LHC Interaction Point (IP2). The proposed Forward Shower Counters (FSCs) are shown as vertical lines (1 to 8). The locations of the dipole (D) and quadrupole (Q) magnet elements are shown as green (dark) and yellow (light) boxes.*

To detect a low-multiplicity decay, a small number (e.g., less than 5) of charged tracks are required in the Central Barrel. Monte Carlo simulation of the detector, beam pipe and magnet elements, has been done with GEANT [7]. It is also to be noted that an average of one interaction per beam crossing is expected with the standard ALICE low luminosity running.

Particle identification by dE/dx for kaons is provided in the range $p < 0.7$ GeV/c. and $p > 1.5$ GeV/c. Efficient tracking is obtained from $p < 100$ MeV/c to $p \approx 100$ GeV/c. A mass resolution of ~ 1% is extimated for two-particle decays in the mass region < 5 GeV. [2].

In summary, the requirement for a trigger would be a low charged multiplicity in the Central Barrel and an absence of a signal in the FMDs and FSCs.[4]

## 3 Central Diffraction (CD) Acceptance

The CD reaction, Eq.(1), was simulated with PHOJET 1.1 [8]. The decay of the central system M into low-multiplicity exclusive final states was generated isotropically with PYTHIA 6.2 [9]. The states $\pi^+\pi^-$, $K^+K^-$, $2\pi^+2\pi^-$ and $K^+K^-\pi^+\pi^-$, are considered for this paper. The central barrel acceptance in pseudo-rapidity corresponds to: $-0.9 < \eta < +0.9$.

The detection efficiencies for events (in the forward direction) were calculated as a function of the diffractive mass, and are shown in Fig. 2. For the final states $\pi^+\pi^-$ and $K^+K^-$, the acceptances range between 4% and 5% ; and for the final states $2(\pi^+\pi^-)$ and $K^+K^-\pi^+\pi^-$, about 2%.

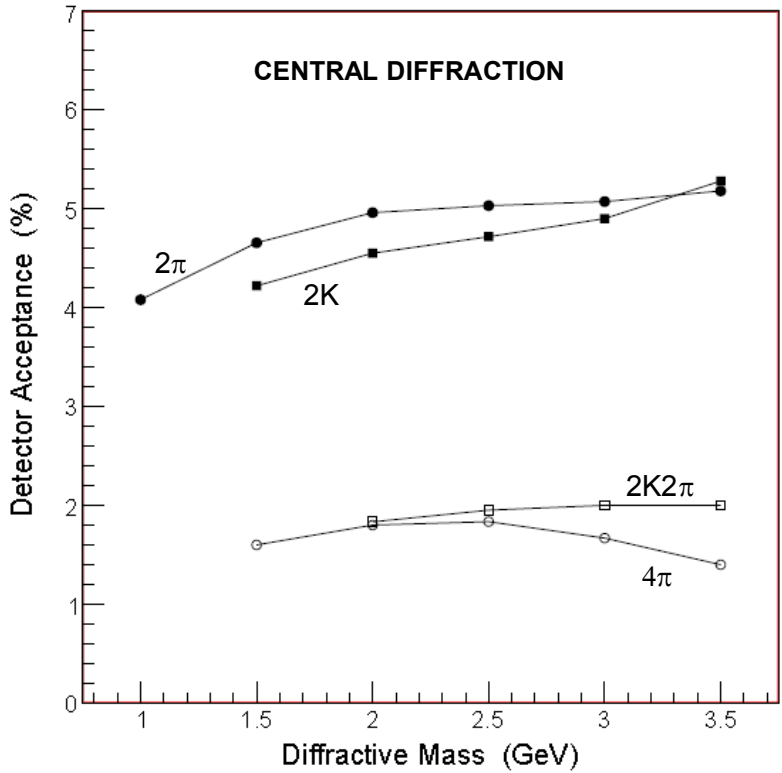

**Figure 2.** *The detector acceptance as a function of the central diffractive mass for $\pi^+\pi^-$, $K^+K^-$, $2\pi^+2\pi^-$ and $K^+K^-\pi^+\pi^-$ decay channels.*

---

[4] Additional conditions could also involve the detection of neutral particles.

## 4 Single Diffraction (SD) Background

Background from single diffraction (SD) is a concern since the multiplicities are mainly forward and will often have a small number of charged particles (less than 5) in the detector acceptance satisfying the Central Barrer requirement. The SD events were generated with PYTHIA. The probability per event that a given number of charged particles fall within the detector acceptance region, as defined above, is shown as the top curve (with filled circles) in Fig.3.

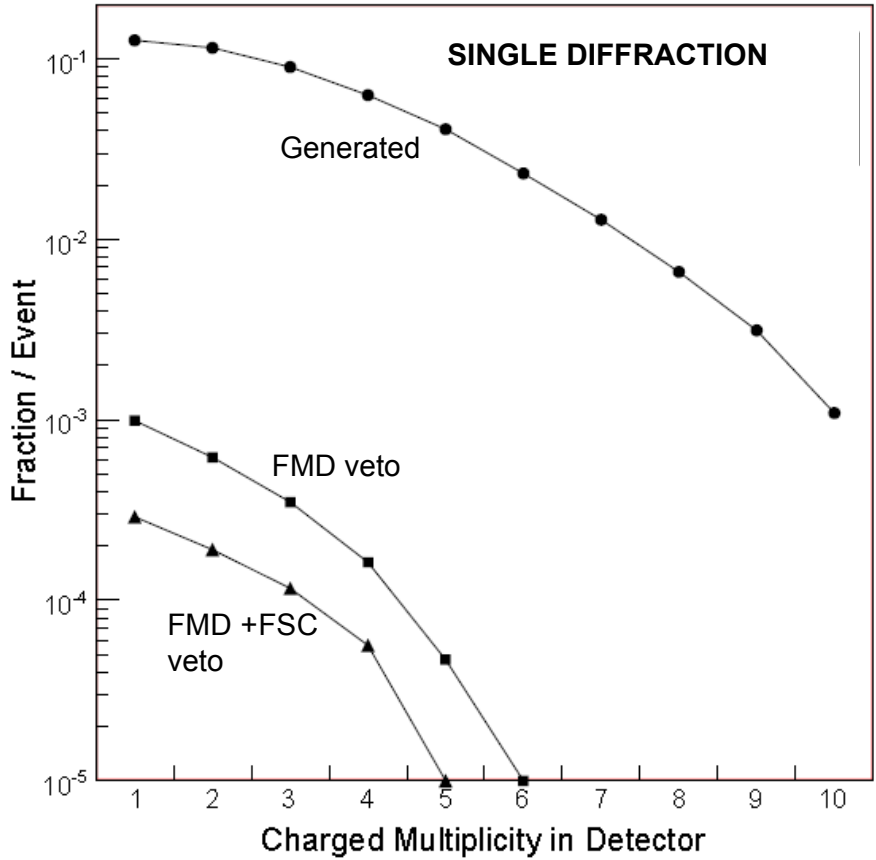

**Figure 3.** *The probability per event for a given number of charged particles to be emitted within the spectrometer detector acceptance region is given by the upper curve (filled circles), the middle curve (filled squares) gives the acceptance with deployment of the FMDs and the lowest curve the combined acceptance of the FMDs and FSCs.*

The value for multiplicities less than five is about 5%. This represents a rather large background which needs to be suppressed. An example of the contribution from this background to the $\pi^+\pi^-$ and $K^+K^-$ mass distributions where a unique $\pi^+\pi^-$ and $K^+K^-$ pair is within the detector acceptance, is shown in Figs.4 and 5.

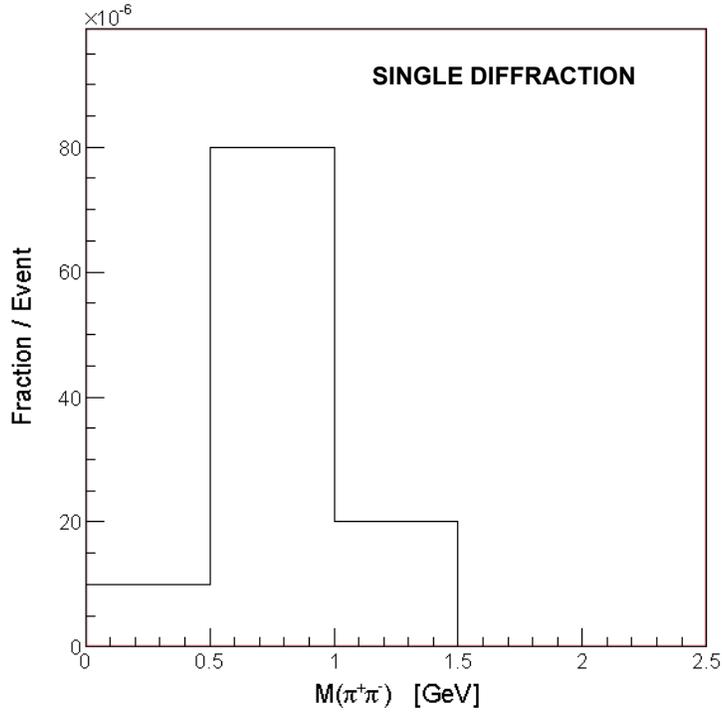

**Figure 4**. *The contribution from single diffractive events that produce a unique $\pi^+\pi^-$ pair within the detector acceptance, as a function of the $\pi^+\pi^-$ effective mass.*

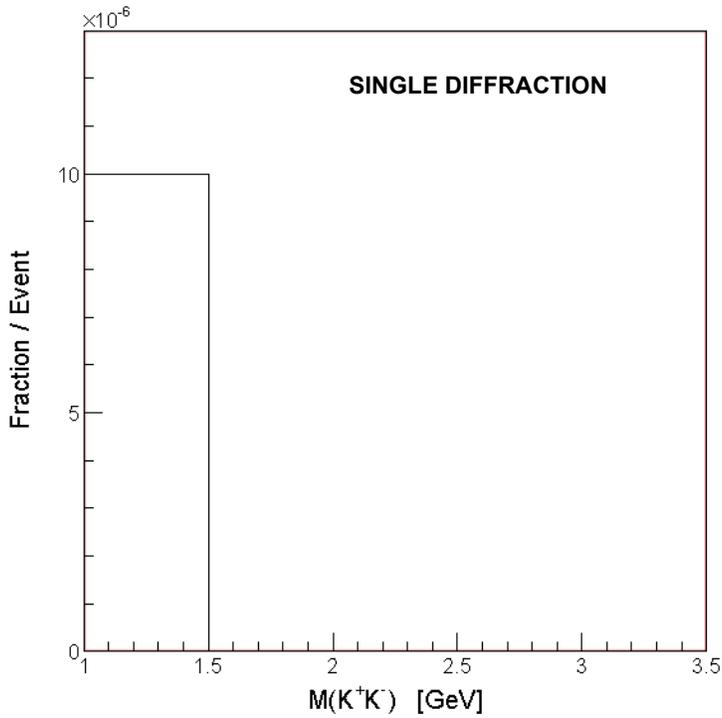

**Figure 5.** *The contribution from single diffractive events that produce a unique $K^+K^-$ pair within the detector acceptance, as a function of the $K^+K^-$ effective mass.*

The deployment of FMDs and FSCs will allow background to be reduced. Single diffraction events will generally produce shower particles from interactions in the beam pipe.[5] The FMD plus FSC detection efficiencies for single diffraction events were calculated as a function of the diffractive mass, see Fig.6. Any track in the FMDs or at least five hits in any of the FSCs has been required. The efficiency is high at the larger masses. Particles from the smallest masses with their low multiplicities rarely enter the ALICE Barrel detector.

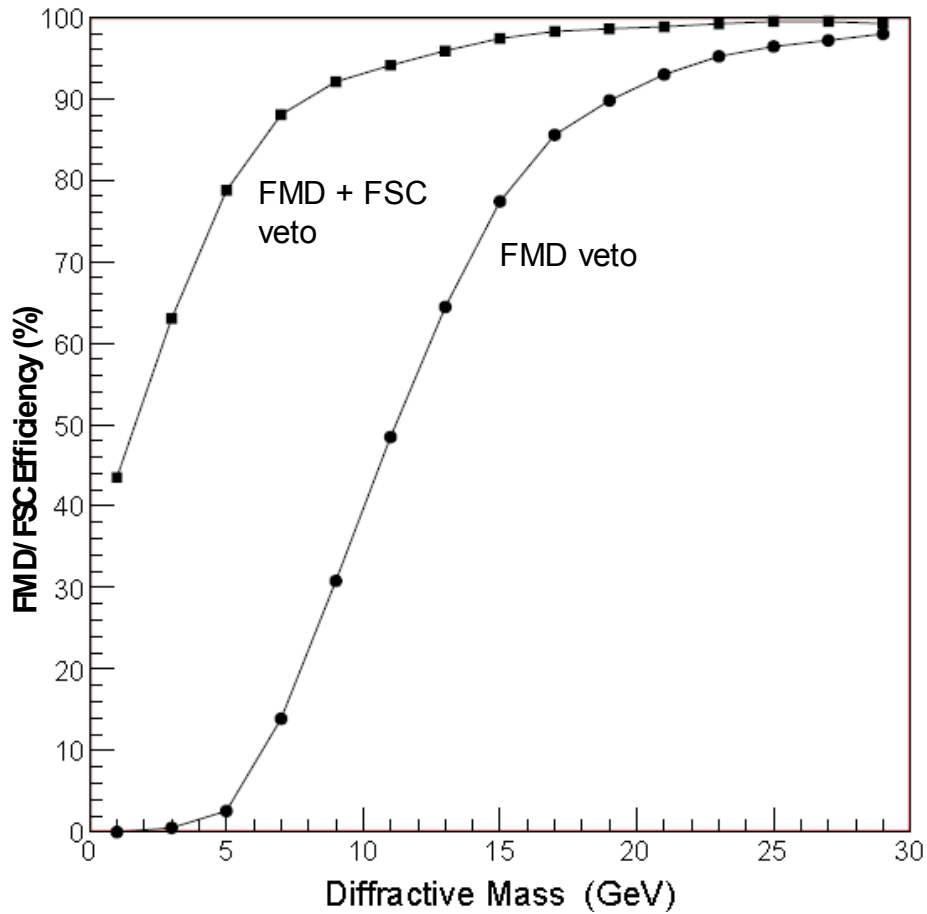

**Figure 6.** *The efficiency to detect single diffractive events (SD) by the FMDs (filled circles) and the combined efficiency of FMDs and FSCs as a function of the diffractive mass.*

With the above FMD plus FSC requirement, the probability per event, that a given number of charged particles enter the detector acceptance region is shown as the bottom curve with filled squares in Fig. 3. The background due to the SD events is reduced by the FMDs plus FSCs by more than two orders of magnitude.[6]

---

[5] The low-mass single diffractive events provide an interesting luminosity monitor, see Ref. [10].
[6] For further discussion, see Ref. [1].

# 5 Non-Diffraction (ND) Background

The analysis is similar to the SD study with the non-diffractive (ND) events generated by PYTHIA. The probability per event, that a given number of charged particles fall within the detector acceptance region, as defined above, is shown in Fig.7. The probability for a small number (less than five) of charged particles is small. This background can be reduced with the FMDs plus FSCs.

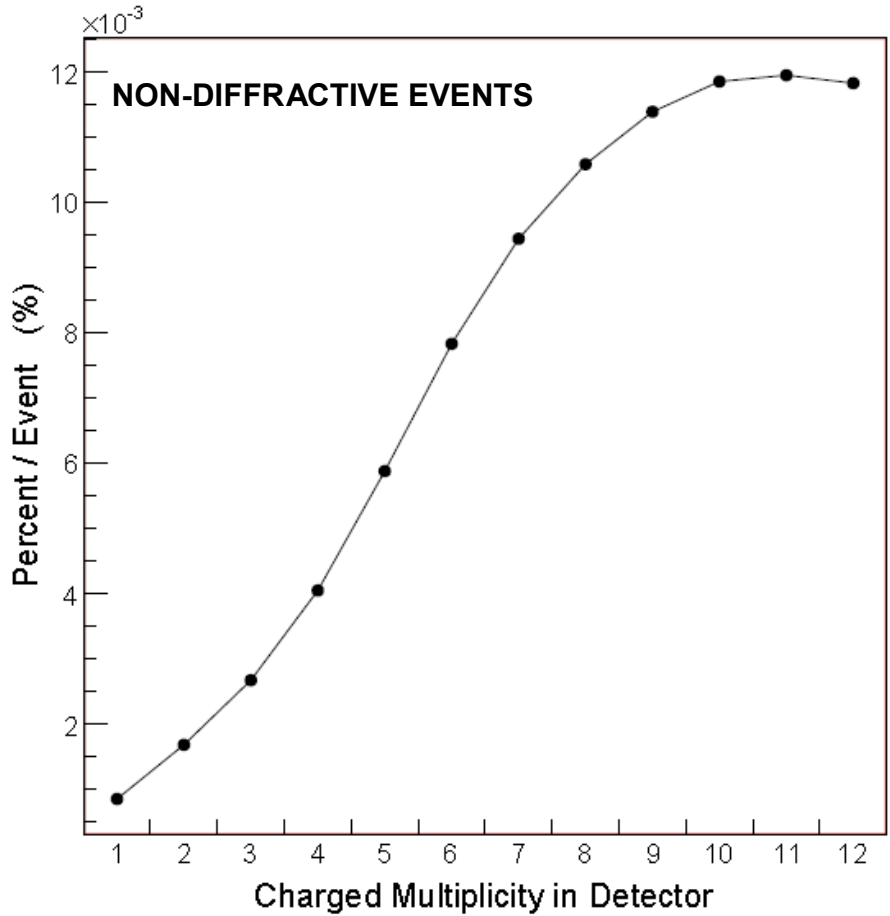

**Figure 7.** *The efficiency to detect non-diffractive events (ND) by the Forward Shower Counters (FSCs) as a function of the charged multiplicity in the detector.*

As in the SD case, the ND events invariably produce shower particles from interactions in the beam pipes. The FMD plus FSC detection efficiencies for the non-diffractive events were calculated as a function of the charged multiplicity in the detector. At least five hits in any of the FMDs/FSCs has been required. The efficiency was essentially 100% over the multiplicity spectrum for the $10^6$ events generated (i.e., all $10^6$ events were detected).

# 6 Central Diffraction Purity

The most important background to the study of the CD exclusive low-multiplicity final states is expected to come from the production and decay of higher mass and multiplicity CD states. The background to the exclusive states resulting from 'feed-down' of higher-mass final-states was calculated with PHOJET. The fraction of cases (purity), where a particle combination, $\pi^+\pi^-$, $K^+K^-$, $2\pi^+2\pi^-$, (within the acceptance) originates from the exclusive decay of the central system M *rather than* from feed-down of higher mass states, is presented in Fig.8.

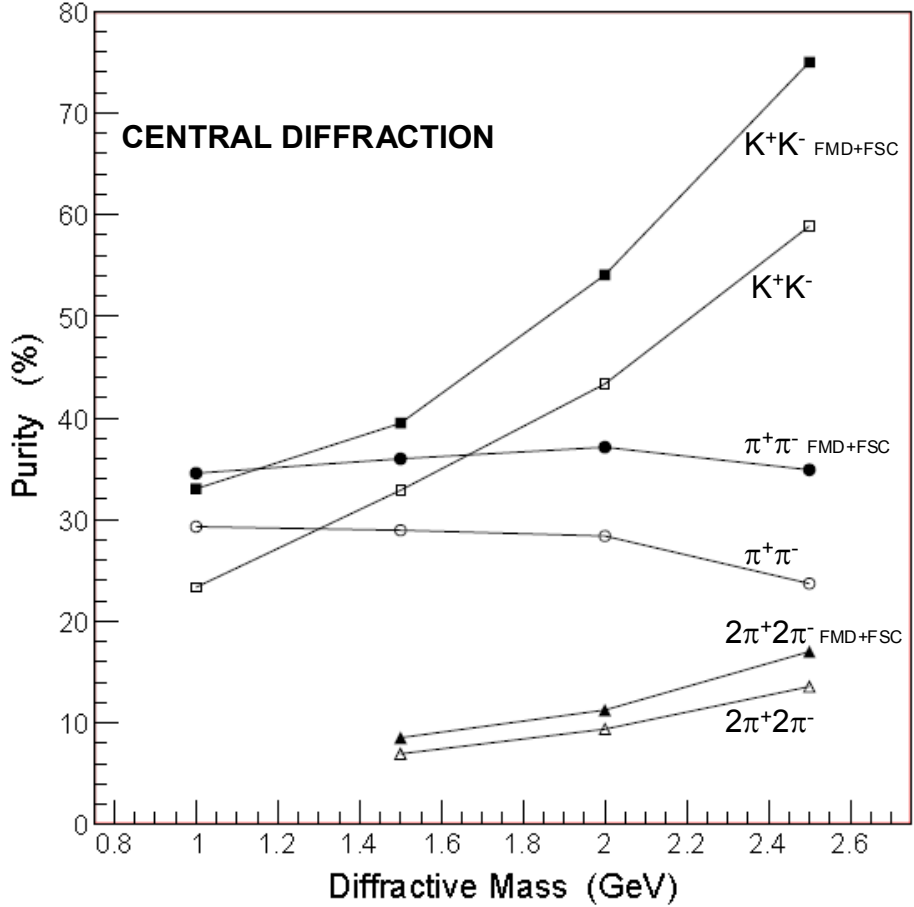

**Figure 8.** *The purities, i.e. the fractions of cases where a particle combination, $\pi^+\pi^-$, $K^+K^-$, $2\pi^+2\pi^-$, (within the acceptance) originates from the exclusive decay of the central system M rather than from feed-down of higher mass states, as a function of the effective mass of the particle combination..*

The purities range from 10% to 40%. The influence of the FMD plus FSC detection efficiencies, for the CD events (by PHOJET) were calculated as a function of the diffractive mass. When the requirement of the presence of rapidity gaps is added, i.e., the purities increase.

# 7 Conclusions

Feasibility studies of the exclusive central diffractive processes for the ALICE experiment have been carried out.[7] With the Forward Multiplicity Detectors (FMDs) and Forward Shower Counters (FSCs), the experiment is shown to be well suited for detailed QCD studies and searches for exotic meson states, such as glueballs, molecules, hybrids and heavy quarkonia.

# Acknowledgements

Invaluable advice from Werner Herr, Valery Khoze and Rainer Schicker, and financial support by the Academy of Finland are gratefully acknowledged.

# References

[1] Jerry W. Lämsä and R. Orava, *Central Diffraction at the LHCb*, Jul 2009. 10pp, JINST 4:P11019,2009, e-Print: arXiv:0907.3847 [physics.acc-ph];Michael Albrow, Albert De Roeck, Valery Khoze, Jerry Lämsä, E. Norbeck, Y. Onel, Risto Orava, and M.G. Ryskin, *Forward physics with the rapidity gaps at the LHC*, arXiv:0811.0120 (2008).

[2] ALICE Technical Design Report,: Forwad detectors, CERN-LHCC-2004-025 ALICE-TDR-011

[3] V.A. Khoze, A.D. Martin, M.G. Ryskin, and W.J. Stirling, *Double-diffractive chi meson production at the hadron colliders*, Eur. Phys. JC35, (2004) 211.

[4] M.G. Ryskin, A.D. Martin, V.A. Khoze and A.G. Shuvaev, *Soft physics at the LHC*, arXiv:0907.1374 (2009), A.D. Martin, M.G. Ryskin, and V.A. Khoze, *Forward Physics at the LHC*, arXiv:0903.2980 [hep-ph], E. Gotsman, E. Levin, U. Maor and J.S. Miller, *A QCD motivated model for soft interactions at high energies*, arXiv:0805.2799 [hep-ph]

[5] Previous results from the CERN ISR: SFM Collaboration, A.Breakstone, et al., Z. Phys. C42 (1989) 387. AFS Collaboration, T.Akesson et al., Nucl. Phys. B 264 (1986) 154. R. Waldi, K.R.Schubert, and K.Winter, Z. Phys. C18 (1983) 301.

[6] CDF results: T. Aaltonen et al., *Observation of exclusive charmonium production and $\gamma\gamma \to \mu^+\mu^-$ in pp collisions at 1.96 TeV*, PRD (2009), T. Aaltonen et al., Phys. Rev. Lett. 98, 112001 (2007), T. Aaltonen et al., Phys. Rev. Lett. 99, 242002 (2007), T. Aaltonen et al**.,** Phys.Rev.D77:052004,2008.

[7] GEANT : R. Brun et al., *GEANT3 Reference Manuel*, DD/EE/84-1, CERN 1987.

[8] PHOJET: R. Engel and J. Ranft, and S. Roesler: *Hard diffraction in hadron-hadron interactions and in photoproduction*, Phys. Rev. D52 (1995) 1459.

[9] PYTHIA: T. Sjöstrand, P. Eden, C. Friberg, L. Lönnblad, G. Miu, S. Mrenna and E. Norrbin, Comput. Phys. Commun. 135, 238 (2001) 238.

[10] V.A. Khoze, A.D. Martin, R. Orava and M.G. Ryskin, *Luminosity Monitors at the LHC*, Eur.Phys. JC19 (2001) 313.

[11] The Helsinki Group: R. Orava with M. Battaglia, O. Bouianov, M. Bouianov, G. Forconi, J. Heino, V. Khoze, A. Kiiskinen, K. Kurvinen, L. Lahtinen, J. Lamsa, V. Nomokonov, A. Numminen, L. Salmi, S. Tapprogge, and M. White with H. Ahola et al. A Technical Report for the ATLAS Collaboration (2000) 93.

---

[7] The authors have carried out an earlier feasibility study of forward physics potential of the ATLAS, CMS and LHCb experiments, see Ref. [11, 1].